\documentclass[
  aps,pre,      
  reprint,        
  superscriptaddress,
  amsmath,amssymb,
  longbibliography,  
  nofootinbib         
]{revtex4-2}

\usepackage{mwe}
\usepackage{graphicx, xcolor, marvosym, lipsum, printlen, float, empheq, tikz, tcolorbox}
\usepackage[caption=false]{subfig}
\usepackage{circuitikz}
\usetikzlibrary{positioning, shapes, calc, arrows.meta, bending, decorations.markings}
\usepackage[caption=false]{subfig}
\usepackage{hyperref}
\usepackage[nameinlink]{cleveref} 
\crefname{appendix}{App.}{Apps.}
\Crefname{appendix}{Appendix}{Appendices}
\crefname{equation}{Eq.}{Eqs.}
\Crefname{equation}{Equation}{Equations}
\crefname{section}{Sec.}{Secs.}
\Crefname{section}{Section}{Sections}
\crefname{figure}{Fig.}{Figs.}
\Crefname{figure}{Figure}{Figures}
\usepackage[normalem]{ulem}
\usepackage{cancel}

\usepackage{mathrsfs, amsmath, amssymb, bm, bbm, mathtools, xfrac, braket, relsize}
\usepackage{comment, verbatim}
\tcbuselibrary{breakable}
\usepackage{minted}
\usepackage[scr=boondox]{mathalpha}
\usepackage{setspace}

\definecolor{fireEngine}{RGB}{193,39,45}
\definecolor{navyBlue}{RGB}{0,0,139}
\definecolor{teal}{RGB}{69,126,136}
\definecolor{caledon}{RGB}{172,234,189}
\definecolor{naplesYellow}{RGB}{246,210,88}

\newcommand{\paren}[1]{\left( #1 \right)}
\newcommand{\sbrac}[1]{\left[ #1 \right]}
\newcommand{\curly}[1]{\left\lbrace #1 \right\rbrace}
\newcommand{\angbr}[1]{\left\langle #1 \right\rangle}
\newcommand{\abs}[1]{\left| #1 \right|}
\newcommand{\e}[1]{\text{e}^{#1}}
\newcommand{\md}{\mathrm{d}}

\newcommand{\approptoinn}[2]{\mathrel{\vcenter{
  \offinterlineskip\halign{\hfil$##$\cr
    #1\propto\cr\noalign{\kern2pt}#1\sim\cr\noalign{\kern-2pt}}}}}

\usepackage[T1]{fontenc}
\usepackage{lmodern}

\newcommand{\appropto}{\mathpalette\approptoinn\relax}

\renewcommand{\sec}[1]{\textit{#1}.}

\date{\today}

\begin{document}

\title{Thermodynamic geometry of friction on graphs: Resistance, commute times, and optimal transport}

\author{Jordan R.\ Sawchuk}
\email{jordan\_sawchuk@sfu.ca}
\author{David A.\ Sivak}
\email{dsivak@sfu.ca}
\affiliation{Department of Physics, Simon Fraser University, Burnaby, British Columbia, Canada V5A1S6}

\date{\today}

\begin{abstract}
    We demonstrate that the thermodynamic friction metric governing dissipation in slowly driven continuous-time Markov chains is equivalent to the commute-time embedding and the resistance distance. This equivalence yields complementary insights: The commute-time embedding demonstrates the intrinsic cost of transporting probability across dynamical bottlenecks, while the resistance distance maps thermodynamic dissipation to Joule heating in an electrical network. We further demonstrate that the linear-response thermodynamic distance is a discrete $L^2$-Wasserstein optimal transport cost evaluated along paths of equilibrium distributions, extending a continuous-state correspondence to discrete networks. This conceptual synthesis of linear-response thermodynamics, random walks on graphs, electrical circuits, and optimal-transport theory connects independently developed geometric frameworks, reduces complex metric calculations to simple circuit algebra, and provides a clear physical picture of dissipation as the energetic cost of routing probability through the state space network.
\end{abstract}

\maketitle
\sec{Introduction}
Geometric ideas have long played a role in thermodynamics, from Riemannian formulations of equilibrium states to geometric treatments of fluctuations, information, and entropy production~\cite{weinhold_metric_1975, salamon_thermodynamic_1983, ruppeiner_riemannian_1995, ito_stochastic_2018, dechantGeometricDecompositionEntropy2022}. In driven stochastic systems, slow control naturally defines a friction metric~\cite{sivakThermodynamicMetricsOptimal2012}. Within this linear-response (LR) regime, the mean excess dissipated power is the squared velocity of the control parameters measured against this metric, and minimum-work control protocols are minimizing geodesics on the thermodynamic manifold. Recently, this framework was connected to optimal-transport (OT) theory~\cite{zhongLinearResponseEquivalence2024}, revealing that for continuous overdamped dynamics, the LR thermodynamic distance coincides with an equilibrium-restricted $L^2$-Wasserstein distance.

Independently, geometries of weighted graphs have emerged in network science~\cite{doyle_random_2000, ghosh2008, spielman_graph_2011, deng_commute_2012, fitch_effective_2019, sato_commute_2019}. In commute-time geometry, the states of a Markov chain are embedded in Euclidean space such that squared distance between states equals the mean round-trip random-walk time~\cite{doyle_commuting_2017}. Closely related is the resistance distance, defined as the effective electrical resistance between nodes in a resistor network constructed on the Markov graph~\cite{chandra_electrical_1989, klein_resistance_1993}. Despite their common dynamical origins, these graph-theoretic geometries have not previously been connected to the thermodynamic geometry of driven processes.

We demonstrate that for discrete continuous-time Markov chains, these geometric frameworks are physically equivalent representations of the same metric structure. This equivalence maps LR dissipation to Joule heating in a resistor network where node potentials are deviations from equilibrium and edge currents are probability fluxes. We exploit this isomorphism to derive exact analytical friction metrics for linear and cyclic graphs. Complementarily, the commute-time embedding provides a local Euclidean description of the thermodynamic manifold, revealing entropic and energetic bottlenecks as distances that are costly to traverse. Finally, we generalize the restricted OT correspondence to discrete networks, framing LR dissipation directly as the energetic cost of routing probability mass through the state space.

\sec{Theoretical background}
We consider a driven, ergodic, continuous-time Markov chain on a finite state space $\Omega$ with $|\Omega| = n+1$. Physically, the state space $\Omega$ typically represents a set of coarse-grained mesostates, such as the set of metastable conformations of a macromolecule. The probability distribution $\bm{p}_s = (p_s(x))_{x\in\Omega}$ evolves according to the master equation 
\begin{equation} \label{eq:masterEqn}
    \frac{1}{\tau_{\text{prot}}}\frac{\partial \bm{p}_{s}}{\partial s} = \mathbb{W}_s\bm{p}_s \ ,
\end{equation}
where $s \in [0,1]$ is the time rescaled by the total protocol duration $\tau_{\text{prot}}$. Control is assumed to be \textit{conservative}, meaning the time-dependence of the transition-rate matrix $\mathbb{W}_s \equiv \mathbb{W}(\bm{V}_s)$ is driven by changing state energies $\bm{V}_s \in \mathbb{R}^{n+1}$ (typically free energies for mesostates). 

We assume that the dynamics are reversible at fixed $\bm{V}_s$, i.e., the transition rates $w_s(x|y)$ (the off-diagonal elements of $\mathbb{W}_s$) satisfy detailed balance $w_{s}(x|y)\pi_s(y) = w_s(y|x)\pi_s(x)$ for instantaneous equilibrium distribution $\pi_s(x) \propto \e{-\beta V_s(x)}$. However, via the response relations derived in~\cite{owen_universal_2020}, the core geometric structures derived here survive relaxation to conservative driving between non-equilibrium steady states (see the Supplemental Material~\cite{supplemental_key}).

Assuming that the system begins in equilibrium at $s=0$, in the quasistatic limit ($\tau_{\text{prot}}\to \infty$) $\bm{p}_s = \bm{\pi}_s$ for all $s$, and the mean dissipated work $\angbr{\mathcal{W}}$ equals the net change $\Delta F$ in free energy. For finite-but-slow driving, the system chases a moving target: a small lag $\delta \bm{p}_s \equiv \bm{p}_s - \bm{\pi}_s$ develops between the actual and instantaneous equilibrium distributions, and this lag produces a mean excess work $\angbr{\mathcal{W}_{\text{ex}}} \equiv \angbr{\mathcal{W}} - \Delta F = \int_0^1\md s\,\delta \bm{p}_s^{\mathsf{T}}\dot{\bm{V}}_s$. To leading order in $\tau_{\text{prot}}^{-1}$, $\delta p$ is linear in protocol velocity and the excess work takes the quadratic form
\begin{equation} \label{eq:lrworkZeta}
    \angbr{ \mathcal{W}_{\text{ex}}}^{\text{LR}} = \frac{1}{\tau_{\text{prot}}}\int_0^1 \md s\,\dot{\bm{V}}_s^{\mathsf{T}} \zeta_{\scriptscriptstyle V_s} \dot{\bm{V}}_s \ .
\end{equation}
The friction tensor $\zeta_{\scriptscriptstyle V} \equiv -\beta\,\mathbb{W}^{\mathcal{D}} D_{\bm{\pi}}$ (for Drazin inverse $\mathbb{W}^{\mathcal{D}}$ of the rate matrix and $D_{\bm{\pi}} \equiv \mathrm{diag}\curly{\bm{\pi}}$) captures the time-integrated relaxation to equilibrium at fixed $\bm{V}$, and via \cref{eq:lrworkZeta} quantifies the energetic cost of motion in the generalized space of discrete energy ``landscapes'' $\bm{V}$~\cite{sivakThermodynamicMetricsOptimal2012, sawchuk_global_2026}. Geometrically, the friction tensor is a metric tensor on the manifold of energy landscapes, and the excess power is proportional to the squared velocity $||\dot{\bm{V}}_s||_{\zeta}^2$ measured in this metric. 

For reversible dynamics, the mapping $\bm{V}\leftrightarrow \bm{\pi}$ is bijective up to a global energy shift. Therefore, the same geometry can be expressed on the space of probability distributions, the (open) probability simplex
\begin{equation}
    \Delta^n = \curly{\bm{p} \in \mathbb{R}^{n+1} \ : \ \bm{1}^{\mathsf{T}}\bm{p} =  1, \, p(x)>0} \ .
\end{equation} 
The metric $g_{\bm{\pi}}$ on $\Delta^n$ is obtained by requiring invariance of excess power under this change of coordinates: $(\md \bm{\pi})^{\mathsf{T}}g_{\bm{\pi}}(\md \bm{\pi}) = (\md \bm{V})^{\mathsf{T}}\zeta_{\scriptscriptstyle V}(\md \bm{V})$. One finds that
\begin{equation} \label{eq:gMat1}
    \beta\,g_{\bm{\pi}} = -D_{\bm{\pi}}^{-1}\mathbb{W}^{\mathcal{D}} \ .  
\end{equation} 
Just as $\zeta_{\scriptscriptstyle V}$ measures a system's resistance to changes in the energy landscape, $g_{\bm{\pi}}$ measures its resistance to changes in the equilibrium distribution. (To avoid notational clutter, we omit the subscripts $s$ and $\bm{\pi}$ through much of this paper.) The right-hand side of~\cref{eq:gMat1} can be read as the Fisher metric $D_{\bm{\pi}}^{-1}$ (the canonical information geometry of the simplex)~\cite{pistoneInformationGeometryProbability2019}, augmented by the relaxation timescales in $-\mathbb{W}^{\mathcal{D}}$; this connection has been noted previously~\cite{sivakThermodynamicMetricsOptimal2012, sawchuk_global_2026}. See S1A in the Supplemental Material [18] for derivations of \eqref{eq:lrworkZeta} and \eqref{eq:gMat1}.

Because total probability is conserved, the simplex $\Delta^n$ for an $(n+1)$-state system is $n$-dimensional, and the tangent space $T_{\bm{\pi}}\Delta^n$ consists of vectors whose elements sum to zero. An $(n+1)\times(n+1)$ representation of a metric on $\Delta^n$ is thus non-unique. We will say that two representations $g$ and $g'$ are \emph{equivalent} on $\Delta^n$ if they define the same length element on the tangent space, written 
\begin{equation} \label{eq:equivDef}
    g_{\bm{\pi}} \overset{\Delta^n}{\sim} g'_{\scriptscriptstyle \bm{\pi}} \iff \dot{\bm{\pi}}^{\mathsf{T}}g_{\bm{\pi}}\,\dot{\bm{\pi}} = \dot{\bm{\pi}}^{\mathsf{T}}g'_{\scriptstyle \bm{\pi}}\,\dot{\bm{\pi}} \ , \ \  \forall \,\bm{\pi}\in \Delta^n, \, \dot{\bm{\pi}} \in T_{\bm{\pi}}\Delta^n \ .
\end{equation}

In practice, control is usually parametric: the equilibrium distribution depends on a lower-dimensional set of experimental parameters $\bm{u}$. The metric~\eqref{eq:gMat1} on the simplex naturally induces a metric on the control-parameter submanifold $\tilde{\zeta}_{ij}(\bm{u}) = \sum_{x,y}g_{\bm{\pi}(\bm{u})}(x,y) \, \partial_i \pi(x) \, \partial_j\pi(y)$~\cite{sawchuk_global_2026} (note that we use subscripts to index partial-control parameters and arguments to index states). Thus all results that follow apply equally to parametric control, with the same equivalence class of metrics governing dissipation.

\sec{Equivalence of linear-response, commute-time, and resistance geometries}
To establish the equivalence of linear-response and graph-theoretic geometries, we associate the Markov chain with a graph $G = (\Omega,\mathcal{E})$ with vertices $x\in \Omega$ and edges $(x,y)\in \mathcal{E}$ connecting states with nonzero transition rates. Under detailed balance, the directed equilibrium flux
\begin{equation} \label{eq:conductance}
    \mathscr{c}(x,y) \equiv w(x|y)\pi(y) = w(y|x)\pi(x) 
\end{equation}
across an edge $(x,y)$ is symmetric. We then define the flux matrix 
\begin{equation} \label{eq:graphLap}
    L(x,y) = \begin{dcases}
        -\mathscr{c}(x,y) & x\neq y \\
        \sum_{z \neq x} \mathscr{c}(x,z) & x= y
    \end{dcases} \ ,
\end{equation} 
or $L = -\mathbb{W}D_{\bm{\pi}}$. For a graph with edge weights $\mathscr{c}(x,y)$, the matrix $L$ is the \textit{weighted graph Laplacian}, ubiquitous in spectral graph theory and (like its continuous namesake) particularly important in the study of diffusion processes on graphs~\cite{mirzaevLaplacianDynamicsGeneral2013, doyle_commuting_2017}. 

The friction metric~\eqref{eq:gMat1} and the Moore-Penrose pseudoinverse $L^+$ \cite{wangGeneralizedInversesTheory2018} of $L$ differ only in their treatment of nonphysical directions corresponding to creation or destruction of probability, and are therefore equivalent on the probability simplex. More precisely, they are related by a projection that shifts their nullspaces:
\begin{equation} \label{eq:glap}
    \beta g = (I - \bm{\pi}\bm{1}^{\mathsf{T}})^{\mathsf{T}} \, L^+ \, (I - \bm{\pi}\bm{1}^{\mathsf{T}}) \ 
\end{equation} 
(see Sec. S1B in the Supplemental Material~\cite{supplemental_key}.)
The projector $I - \bm{\pi}\bm{1}^{\mathsf{T}}$ acts as the identity on all admissible (probability-conserving) $\dot{\bm{\pi}}$, so \cref{eq:glap} immediately implies that $\beta g \overset{\Delta^n}{\sim} L^+$. 

We can map the Markov chain to a resistor network by defining the resistance of an edge as the inverse of the directed equilibrium flux. Then $\mathscr{c}$ is a conductance and $L$ is the network's admittance matrix. For node current injections $\dot{\pi}(x)$, by Ohm and Kirchhoff's laws (elaborated in \textit{Linear-response dissipation as Joule heating}) $L^+\dot{\bm{\pi}}$ is a vector of node potentials. The effective resistance between nodes $x, y \in \Omega$ is
\begin{subequations} \label{eq:effRes}
   \begin{align}
    R_{\text{eff}}(x,y) &= (\hat{\mathbf{e}}_x - \hat{\mathbf{e}}_y)^{\mathsf{T}} \, L^+ \, (\hat{\mathbf{e}}_x - \hat{\mathbf{e}}_y) \\ 
    &= L^+(x,x) + L^+(y,y) - 2L^+(x,y) \ , 
    \end{align}
\end{subequations}
for unit basis vectors $\hat{\mathbf{e}}_x, \hat{\mathbf{e}}_y \in \mathbb{R}^{|\Omega|}$~\cite{klein_resistance_1993, yadav_network_2015}. It follows immediately from the definition of the tangent space that $R_{\text{eff}} \overset{\Delta^n}{\sim}-2L^+$, and thus $\beta \,g \overset{\Delta^n}{\sim} -\tfrac{1}{2}R_{\text{eff}}$.

Finally, the mean commute time $C(x,y)$ between states $x$ and $y$ is defined as the average time to travel from $y$ to $x$ and back again (or vice versa),
\begin{equation} \label{eq:commtime}
    C(x,y) \equiv \tau_{\rm{mfp}}(x|y) + \tau_{\rm{mfp}}(y|x) \ ,
\end{equation}
with $\tau_{\rm{mfp}}(x|y) \equiv \angbr{\inf \curly{t : X_t = x} \ | \ X_0 = y}$ the mean first-passage time (MFPT) from $y$ to $x$. Using the relation
\begin{equation} \label{eq:DrazinMFPT}
    \mathbb{W}^{\mathcal{D}} = D_{\bm{\pi}} \, \mathcal{T}_{\text{mfp}} \, (I - \bm{\pi}\bm{1}^{\mathsf{T}}) \   
\end{equation}
between the Drazin inverse of the rate matrix and the MFPTs~\cite{coolen-schrijnerDeviationMatrixContinuoustime2002} [here $\mathcal{T}_{\text{mfp}}$ is the matrix whose $x,y$ component is $\tau_{\rm{mfp}}(x|y)$], direct substitution into \cref{eq:gMat1} yields $\beta\,g  \overset{\Delta^n}{\sim} -\mathcal{T}_{\text{mfp}}$ after dropping the projector $I-\bm{\pi}\bm{1}^{\mathsf{T}}$ as before. Since the LR dissipation is governed by a quadratic form (physically, this reflects the time-reversal symmetry of the lowest-order approximation of the excess work), only the symmetric part of the MFPT matrix contributes, so $\beta g \overset{\Delta^n}{\sim}-\tfrac{1}{2}C$ with $C \equiv \mathcal{T}_{\text{mfp}} + \mathcal{T}_{\text{mfp}}^{\mathsf{T}}$. 

To summarize, we have shown the following:
\begin{equation} \label{eq:centralEquivs}
    \beta g \overset{\Delta^n}{\sim} L^+ \overset{\Delta^n}{\sim} -\tfrac{1}{2}R_{\text{eff}} \overset{\Delta^n}{\sim} -\tfrac{1}{2}C \ .
\end{equation}
These metric equivalences constitute a central result of this paper. The LR thermodynamic, resistance, and commute-time geometries---all unified by the graph Laplacian---are different manifestations of the same network structure. Moreover, these four matrices uniquely determine one another (see Sec. S2 in the Supplemental Material~\cite{supplemental_key}). The implications of these equivalences are explored below.

\sec{Thermodynamic distance and optimal transport on graphs}
Recent work has utilized $L^1$ optimal-transport costs to establish thermodynamic speed limits in discrete systems~\cite{van_vu_thermodynamic_2023}. Here, we show a complementary correspondence with $L^2$-Wasserstein OT, extending known results from continuous overdamped dynamics~\cite{zhongLinearResponseEquivalence2024}: The squared thermodynamic distance
\begin{equation}
    \mathcal{L}^2(\bm{\pi}_0,\bm{\pi}_1) \equiv \inf_{\bm{\pi}_s}\int_0^1 \md s \,\,\dot{\bm{\pi}}_s^{\mathsf{T}}(\beta g) \,\dot{\bm{\pi}}_s
\end{equation}
between equilibrium distributions $\bm{\pi}_0, \bm{\pi}_1 \in \Delta^n$ equals a discrete $L^2$-Wasserstein transport cost evaluated along paths of equilibrium distributions.

For two continuous densities $\rho_0$, $\rho_1$ on $\mathbb{R}^{d}$, the Benamou-Brenier formulation~\cite{benamou_computational_2000, otto_geometry_2001} of the $L^2$-Wasserstein distance is 
\begin{equation}\label{eq:BBD}
        \mathcal{W}_2^2(\rho_0,\rho_1) = \inf_{\dot{\rho}_s =- \nabla \cdot(\rho_s\nabla\phi_s)} \,  \int_0^1\md s \int \md x\,\rho_s(x)|\nabla\phi_s(x)|^2 \ ,
\end{equation} 
where $\phi_s$ is the velocity potential. Translating this continuous picture to a discrete network, we map continuous vector fields to edge fluxes and scalar fields to node potentials, using techniques from discrete calculus~\cite{grady_discrete_2010}. The analog of the continuity equation $\dot{\rho}_s = - \nabla\cdot(\rho_s\nabla \phi_s)$ that emerges from this mapping is
\begin{equation} \label{eq:nodePots}
    \dot{\bm{\pi}}_s = L_s \phi_s \ ,
\end{equation}
with $\phi_s\in\mathbb{R}^{|\Omega|}$ (defined up to additive constant) and $L_s$ respectively acting as velocity potential and weighted Laplacian [$-\nabla\cdot(\rho_s\nabla\cdot)$]. The probabilistic interpretation of the potential $\phi_s$ is elaborated in \textit{Node potentials and edge currents}.

Geometrically, the $\phi_s$ are covectors, and the graph Laplacian $L_s$ is the cometric of the friction tensor: $\dot{\bm{\pi}}_s^{\mathsf{T}}L^+\dot{\bm{\pi}}_s =\phi_s^{\mathsf{T}}\dot{\bm{\pi}}_s= \phi_s^{\mathsf{T}}L_s \phi_s$. With graph gradient $[\nabla_G \phi_s](x,y) = \phi_s(x) - \phi_s(y)$ for $(x,y) \in\mathcal{E}$ and conductance-weighted norm $||\cdot||_{\mathcal{H}(\mathcal{E})}$ on the edge space (see Sec. S4A in the Supplemental Material~\cite{supplemental_key} for formal definitions), we have $\phi_s^{\mathsf{T}}L_s\phi_s = ||\nabla_G\phi_s||^2_{\mathcal{H}(\mathcal{E})}$. The squared thermodynamic distance between two equilibrium distributions $\bm{\pi}_0, \bm{\pi}_1$ is then
\begin{equation}
        \mathcal{L}^2(\bm{\pi}_0,\bm{\pi}_1) = \inf_{\dot{\bm{\pi}}_s=L_s\phi_s} 
        \int_0^1 \md s \abs{\abs{\nabla_G \phi_s}}_{\mathcal{H}(\mathcal{E})}^2 \ . \label{eq:equivOT}
\end{equation}
\Cref{eq:equivOT} is an equilibrium-path-restricted variant of the discrete $L^2$-Wasserstein metric introduced in \cite{maas_gradient_2011, chowFokkerPlanckEquations2012} (see Sec. S4B in the Supplemental Material~\cite{supplemental_key}). There are some formal differences between the continuous~\eqref{eq:BBD} and discrete~\eqref{eq:equivOT} expressions: the equilibrium weights are absorbed into the definition of the edge-flux inner product and the graph Laplacian in the discrete case. However, both expressions describe a quadratic instantaneous dissipative cost associated with probability currents driven by a potential field, subject to a mass conservation equation. The connection between discrete OT and the graph Laplacian was also noted in~\cite{liTransportInformationGeometry2022}.

\sec{Commute-time embedding and bottlenecks} 
Above, we showed that the friction metric $g$ and the commute-time matrix $C$ encode the same geometry on the probability simplex $\Delta^n$. A classical result states that $C$ is a squared Euclidean distance matrix~\cite{doyle_commuting_2017}: there exists an embedding $\Omega \ni x \mapsto \bm{a}(x) \in\mathbb{R}^m$ with $m\leq n=|\Omega|-1$ such that 
\begin{equation} \label{eq:commuteEmb}
    C(x,y) = ||\bm{a}(x) - \bm{a}(y)||^2 \ .
\end{equation}
Through this embedding, the Markov graph---a purely topological construction---acquires a geometry in which each state $x$ sits at a point $\bm{a}(x) \in \mathbb{R}^m$. 

To illustrate the physical significance of this embedding, consider transferring a small amount $\epsilon$ of probability mass from state $y$ to state $x$. The required work in linear response is simply
\begin{equation}
        \angbr{\md \mathcal{W}_{\text{ex}}}^{\text{LR}} = \epsilon^2 \,k_{\rm B} T\,||\bm{a}(x) - \bm{a}(y)||^2 \ . 
\end{equation}
That is, the linear-response cost of transporting probability between the two states is quadratic in the distance between them in the commute-time embedding. For general $\md \bm{\pi}$, the work increment is
\begin{equation} \label{eq:workInc}
    \angbr{\md \mathcal{W}_{\text{ex}}}^{\text{LR}} = k_{\rm B} T\,||A^{\mathsf{T}}\md\bm{\pi}||^2 \ ,
\end{equation}
where the matrix $A$ encodes the embedded positions of the states [$A(x,\cdot)=\bm{a}(x)$] and may be obtained from $C$ via classical multidimensional scaling~\cite{wang_classical_2012}. \Cref{eq:workInc} admits a centroid interpretation: states with positive (negative) increments define a weighted centroid of probability-increasing (probability-decreasing) states in the Euclidean embedding, and the cost of transport is the squared distance between these centroids. 

Geometrically, the commute-time embedding provides a flat local map of the thermodynamic manifold, with the dissipative cost of a small step $\md \bm{\pi}$ behaving like a Euclidean distance $\md \ell = \sqrt{(\md x_0)^2 + (\md x_1)^2 + \cdots + (\md x_n)^2}$ in the coordinates $\md x_i = (A^{\mathsf{T}}\md \bm{\pi})(x_i)$.

The commute-time embedding also reveals bottlenecks in the dynamics. Sets of states with relatively short pairwise commute times form \emph{clusters} in the embedding, and large gaps between clusters are bottlenecks. \Cref{eq:workInc} says that transporting probability mass between clusters is expensive, while redistributing mass within a cluster is cheap. 

We mark two distinct origins for such bottlenecks, which we refer to as energetic and entropic bottlenecks, borrowing terminology from molecular kinetics~\cite{zwanzig_dynamical_1992, chakrabarti_waiting_2003}. Energetic bottlenecks occur when the allowed paths between two regions involve at least one intermediate state with a large energy, creating long relaxation times and thus large commute distances between the regions. These originate in the potential landscape rather than the network topology, and can often be mitigated by control parameters that lower relative barrier heights. Entropic bottlenecks, on the other hand, arise when few transition pathways connect two clusters of states: Even when inter-cluster rates are comparable to intra-cluster rates, a sparse connectivity forces trajectories through narrow channels. Such bottlenecks are topological and cannot be removed by conservative control, so there is an unavoidable cost of moving probability between clusters separated by an entropic bottleneck.  

\sec{Linear-response dissipation as Joule heating}
The equivalence $\beta g \overset{\Delta^n}{\sim}-\tfrac{1}{2}R_{\text{eff}}$ gives a complementary physical picture: each edge $(x,y)  \in \mathcal{E}$ acts as a branch with a resistance $r(x,y) = [w(x|y)\pi(y)]^{-1}$ under detailed balance. The discrete continuity equation $\dot{\bm{\pi}} = L\phi$ introduced in \textit{Thermodynamic distance and optimal transport on graphs} may then be written as
\begin{equation} \label{eq:KCL}
    \dot{\pi}(x) = \sum_{y}i(x,y) 
\end{equation}
for edge currents (directed from $y$ to $x$)
\begin{equation} \label{eq:currents}
    i(x,y) = \frac{\phi(x) - \phi(y)}{r(x,y)} \ .
\end{equation}
\Cref{eq:currents} is Ohm's law for node potentials $-\phi(x)$ and edge currents $i(x,y)$, and Eq.~\eqref{eq:KCL} is Kirchhoff's current law with node current injections $\dot{\pi}(x)$. The linear-response excess work is then
\begin{equation} \label{eq:circuitWork}
    \braket{\mathcal{W}_{\text{ex}}}^{\text{LR}} = \frac{k_{\rm B} T}{\tau_{\text{prot}}}\int_0^1\md s\sum_{(x,y) \in \mathcal{E}} r_s(x,y)\,i_s(x,y)^2 \ .    
\end{equation}
The integrand has the exact mathematical form of the power dissipated in a resistor network: driving probability currents $i_s(e)$ along the edges $e \in \mathcal{E}$ incurs a quadratic cost governed by the instantaneous edge resistances $r_s(e)$. Geometrically, \cref{eq:circuitWork} tells us that the friction metric is globally diagonalized when expressed on the $|\mathcal{E}|$-dimensional edge space. We leverage this simplification to derive exact results in \textit{Metrics for elementary topologies}.

\sec{Node potentials and edge currents}
The scalar field $\phi$ now appears (up to sign convention) as both the electrical potential generating edge currents in the resistor network and the velocity potential generating probability fluxes in the discrete OT formulation. We now provide a more direct probabilistic interpretation of $\phi(x)$, and in doing so, we clarify the nature of the edge currents $i(x,y)$. 

The linear-response approximation of the lag $\delta \bm{p}$ implicit in the friction-tensor formalism is \cite{sawchuk_global_2026, mandalAnalysisSlowTransitions2016, avron_adiabatic_2012} 
\begin{equation} \label{eq:pDev}
    \delta \bm{p}_s \approx \delta \bm{p}^{\text{LR}}_s = \frac{1}{\tau_{\text{prot}}}\,\mathbb{W}^{\mathcal{D}}_s\dot{\bm{\pi}}_s  \ ,
\end{equation}
valid for sufficiently long $\tau_{\text{prot}}$.
A constant offset of $\phi$ makes no physical difference, so taking the gauge $\braket{\phi}_{\pi} = 0$ without loss of generality, combining \cref{eq:nodePots,eq:pDev} yields
\begin{equation} 
    -\phi(x) = \tau_{\text{prot}}\frac{\delta p^{\text{LR}}(x)}{\pi(x)}  \ . 
\end{equation}
Note that since the lag $\delta p^{\text{LR}}(x)$ is $\mathcal{O}(\tau_{\text{prot}}^{-1})$, $\phi(x)$ is $\mathcal{O}(1)$ in $\tau_{\text{prot}}$. The electric potential $-\phi(x)$ physically represents the excess probability at $x$ relative to the equilibrium distribution. Substituting this into \eqref{eq:currents} and applying detailed balance gives 
\begin{equation}
    \frac{i(x,y)}{\tau_{\text{prot}}} = w(x|y) \,p^{\text{LR}}(y) - w(y|x)\, p^{\text{LR}}(x) \ .
\end{equation}
The edge currents $i(x,y)$ [like the potentials, $\mathcal{O}(1)$ in $\tau_{\text{prot}}$] are precisely the (unitless) probability currents in the linear-response regime, due to relaxation of the small deviation from equilibrium quantified by the node potentials $\phi(x)$. 

We note a structural similarity to a circuit mapping derived for systems subject to time-constant nonconservative forces~\cite{linCircuitReductionHeterogeneous2020}. In that work, resistors, potentials, and currents are defined identically to the derived quantities presented here, and the results are leveraged to obtain stationary fluxes, generalized reciprocal relations, and amplification bounds far from equilibrium. The shared mathematical foundation suggests an intriguing avenue for extending this simple geometric formalism beyond linear response.

\begin{figure}[t]
    \resizebox{\linewidth}{!}{
    \begin{tikzpicture}
        [
        node_style/.style = {circle, draw=black, thick, minimum size=0.6cm, fill=white, inner sep=0pt, font=\bfseries},
        elec_node/.style = {circle, draw, fill=black, minimum size=3pt, inner sep=0pt},
        r/.style = {start angle=130, end angle=50, x radius=0.75cm, y radius=0.6cm},
        l/.style = {start angle=310, end angle=230, x radius=0.75cm, y radius=0.6cm},
        circuitikz/bipoles/length=0.8cm,
        circuitikz/resistors/scale=1,
        ]
        \def\limit{4.5}
        \path (-\limit, 0) -- (\limit, 0);

        \node[anchor=west] (label_a) at (-\limit, 2.5) {(a)};
        \node[anchor=west] (label_b) at (-\limit, -1.4) {(b)};

        \begin{scope}[shift={(-0.25,0)}]
        
        \def\offset{1.7}
        \node[node_style, fill=caledon!20] (0) at (-3, \offset) {$0$};
        \node[node_style, fill=caledon!20] (1) at (-1.5, \offset) {$1$};
        \node[node_style, fill=caledon!20] (2) at (0, \offset) {$2$};
        \node[] (blank) at (1.5, \offset) {};
        \node[] (blank2) at (2.05, \offset) {};
        \node[node_style, fill=caledon!20] (n) at (3.5,\offset) {$n$};
        \draw (blank) -- (blank2) [dotted, very thick];
        \draw[-latex] (0.north east) arc[r] node[midway, above]{$\scriptstyle w(1|0)$};
        \draw[-latex] (1.south west) arc[l] node[midway, below]{$\scriptstyle  w(0|1)$};
        \draw[-latex] (1.north east) arc[r] node[midway, above]{$\scriptstyle w(2|1)$};
        \draw[-latex] (2.south west) arc[l] node[midway, below]{$\scriptstyle  w(1|2)$};        
        \draw[-latex] (2.north east) arc[r] node[midway, above]{$\scriptstyle w(3|2)$};
        \draw[-latex] ($(blank.south west) - (0.1,0.1)$) arc[l] node[midway, below]{$\scriptstyle  w(2|3)$};   
        \draw[-latex] (blank2.north east) arc[r] node[midway, above]{$\scriptstyle w(n|n-1)$};
        \draw[-latex] (n.south west) arc[l] node[midway, below]{$\scriptstyle  w(n-1|n)$};   

        \node[elec_node] (0) at (-3,0){};   \node[anchor=south] () at (-3,0.05) {$0$};
        \node[elec_node] (1) at (-1.5,0){};   \node[anchor=south] () at (-1.5,0.05) {$1$};
        \node[elec_node] (2) at (0,0){};   \node[anchor=south] () at (0,0.05) {$2$};
        \node[elec_node] (n) at (3.5,0){};   \node[anchor=south] () at (3.5,0.05) {$n$};
        \node[] (blank) at (1.5, 0) {};
        \node[] (blank2) at (2.05,0) {};
        \draw (0) to[R, name=r0] (1); \node[above=2pt of r0.north] {$\scriptstyle  r(0)$};
        \draw (1) to[R, name=r1] (2); \node[above=2pt of r1.north] {$\scriptstyle r(1)$};
        \draw (2) to[R, name=r2] (blank); \node[above=2pt of r2.north] {$\scriptstyle r(2)$};
        \draw (blank) -- (blank2) [dotted, very thick];
        \draw (blank2) to[R, name=rlst] (n); \node[above=2pt of rlst.north] {$\scriptstyle r(n-1)$};
        \draw[->, thick, fireEngine] (-1.8,-0.4) -- (-2.7,-0.4);
        \node[fireEngine, below=10pt of r0.south] {$\scriptstyle i_0(0)$};
        \draw[->, thick, fireEngine] (-0.3,-0.4) -- (-1.2,-0.4);
        \node[fireEngine, below=10pt of r1.south] {$\scriptstyle i_0(1)$};
        \draw[->, thick, fireEngine] (1.2,-0.4) -- (0.3,-0.4);
        \node[fireEngine, below=10pt of r2.south] {$\scriptstyle i_0(2)$};
        \draw[->, thick, fireEngine] (3.2,-0.4) -- (2.35,-0.4);
        \node[fireEngine, below=10pt of rlst.south] {$\scriptstyle i_0(n-1)$};
        \end{scope}
        
        \begin{scope}
        [
        shift={(-0.1,-3.8)},
        linear_current/.style={->,>=Latex,color=fireEngine, line width = 1},
        cycle_current/.style={->, >=Latex, color=navyBlue, line width = 1},
        circuitikz/resistors/scale=1
        ]
            \def\R{2.0}
    
            \node[elec_node](n-1)at(190:\R){}; \node[]()at(190:\R+0.6){$n-1$};
            \node[elec_node](n)at(145:\R){};\node[]()at(145:\R+0.4){$n$};
            \node[elec_node](0)at(100:\R){};\node[]()at(100:\R+0.4){$0$};
            \node[elec_node](1)at(55:\R){};\node[]()at(55:\R+0.4){$1$};
            \node[elec_node](2)at(10:\R){};\node[]()at(10:\R+0.4){$2$};
            \node[elec_node](3)at(-35:\R){};\node[]()at(-35:\R+0.4){$3$};
    
            \draw[bend left](-35:\R)arc(-35:-60:\R);
            \draw[bend right](190:\R)arc(190:215:\R);
            \draw[dotted, very thick, bend right] (-90:\R) arc (-90:-115:\R);
    
            \draw[bend right](n-1)arc(190:180:\R);
            \draw (180:\R) to[R, name=rn-1] (155:\R);
            \draw[bend right](155:\R)arc(155:145:\R);
    
            \draw[bend right](n)arc(145:135:\R);
            \draw (135:\R) to[R, name=rn] (110:\R);
            \draw[bend right](110:\R)arc(110:100:\R);
    
            \draw[bend right](0)arc(100:90:\R);
            \draw (90:\R) to[R, name=r0] (65:\R);
            \draw[bend right](65:\R)arc(65:55:\R);
    
            \draw[bend right](1)arc(55:45:\R);
            \draw (45:\R) to[R, name=r1] (20:\R);
            \draw[bend right](20:\R)arc(20:10:\R);
    
            \draw[bend right](2)arc(10:0:\R);
            \draw (0:\R) to[R, name=r2] (-25:\R);
            \draw[bend right](-25:\R)arc(-25:-35:\R);
    
            \draw[cycle_current](-260:0.7)arc(-260:140:0.8) node[midway, above] {$\scriptstyle i_{\text{cyc}} \ \ $};
            \draw[cycle_current, color=fireEngine,|->|] (-210:1.2) arc (-210:95:1.2) node[midway, below] {$\scriptstyle \ \ \ \ i_0(x)$};
        \end{scope}
        
    \end{tikzpicture}
    }
    \caption{(a) The Markov graph for a linear chain of states and its series circuit representation. The orientation of the edge currents $i_0$ reflects the convention in the main text. (b) Circuit representation of a cyclic Markov graph, with total currents $i(x)$ decomposed into a reference current $i_0(x)$ obtained by a cut along $(n,0)$ and a cycle correction $i_{\text{cyc}}$.}
    \label{fig:graphs}
\end{figure}

\sec{Metrics for elementary topologies}
By treating the Markov graph as a physical circuit, we can bypass complex matrix inversions and use standard tools like series/parallel reduction and Kron reduction~\cite{dorfler_kron_2013} to directly compute the friction metric. For simple topologies, we may instead derive closed-form expressions for the currents and make use of \cref{eq:circuitWork}.

We denote by $\mathcal{P} \equiv \sum_{(x,y)\in\mathcal{E}}r_s(x,y)\, i_s(x,y)^2$ the ``power'' in the resistor network model, equal (up to dimensional factors) to the physical LR excess power. In this section, we derive closed-form expressions for $\mathcal{P}$ for linear and cyclic graphs. These expressions give the excess work of any protocol explicitly in terms of edge resistances, bypassing the need for matrix inversion.

\sec{Linear graph}
Consider a chain of $n+1$ states connected by $n$ edges $(x,x+1)$. We label edges by the lower node value as in \hyperref[fig:graphs]{Fig.$\,$1a}, and we denote edge currents by $i_0(x)$ (adding the subscript 0 in anticipation of their role as a reference current for the cyclic graph). Because there are no loops, the continuity equation \eqref{eq:KCL} can be inverted as $i_0(x) = \sum_{y=0}^x \dot{\pi}(y) \equiv \dot{\Pi}(x)$ for equilibrium cumulative distribution function $\Pi(x)$. For an arbitrary set of $m$ control parameters $\bm{u} = \curly{u^0,u^1,\dots,u^{m-1}}$ we have (with Einstein summation over parameter indices) $\dot{\Pi}(x) = \partial_i \Pi(x) \, \dot{u}^i$, so
\begin{equation} \label{eq:linPow} 
    \mathcal{P}_{\text{lin}} = \underbrace{\sum_{x=0}^{n-1} \frac{\partial_i \Pi(x) \, \partial_j \Pi(x)}{\mathscr{c}(x)}}_{\propto \tilde{\zeta}_{ij}^{\text{lin}}}\,\dot{u}^i\dot{u}^j \ ,
\end{equation}
from which we immediately identify the partial-control friction metric $\tilde{\zeta}_{ij}^{\text{lin}}$. This is the discrete analog of the friction tensor $\tilde{\zeta}_{ij} = \int \md x \, \frac{\partial_i\Pi(x)\, \partial_j\Pi(x)}{D\pi(x)}$ for 1D overdamped Langevin dynamics~\cite{zulkowskiOptimalControlOverdamped2015}, with $\sum \to \int \md x$ and $\mathscr{c}(x) \to D \pi(x)$. This aligns (up to dimensional factors) with recent work showing that the symmetrized flux across an edge [exactly $\mathscr{c}(x)$ under detailed balance] becomes $\beta D \pi(x)$ in the continuous limit~\cite{van_vu_thermodynamic_2023}.

One consequence of~\cref{eq:linPow} is that along any LR minimum-work protocol for any linear chain, $\sum_{x=0}^{n-1}\sbrac{w_s(x+1\,|\,x)\, \pi_s(x)}^{-1}\dot{\Pi}_s(x)^2$ is a constant of motion, given entirely in terms of elementary quantities. In particular, if the total dissipation is dominated by a single high-resistance edge $x$, then in the optimal strategy $\dot{\Pi}_s(x) \appropto [r_s(x)]^{-1/2}$.

\sec{Cycle graph}
A cycle graph (\hyperref[fig:graphs]{Fig.$\,$1b}) is formed by adding a single edge $(n,0)$ to the linear graph. We decompose the true currents $i(x)$ [with convention $i(x) = i(x,x+1)$ for summation modulo $n+1$] into a reference current and a cycle correction:
\begin{equation}
    i(x) = i_0(x) + i_{\text{cyc}}(x) \ .
\end{equation} 
Here, $i_0(x) = \dot{\Pi}(x)$ is the current that would flow under the same driving if the loop were cut at $(n,0)$. Because $i_0$ satisfies the inhomogeneous Kirchhoff's current law~\eqref{eq:KCL}, the correction $i_{\text{cyc}}$ must satisfy $\sum_{y}i_{\text{cyc}}(x,y) = 0$ for all $x$, meaning it is a spatially uniform loop current. In particular, $i_{\text{cyc}} = i(n)$, the current on the cut edge $(n,0)$.

The magnitude of $i_{\text{cyc}}$ can be determined by Thomson's principle~\cite{doyle_random_2000}: the currents $i(x)$ are those that uniquely minimize $\mathcal{P}$ subject to Kirchhoff's current law [\cref{eq:KCL}]. We obtain
\begin{equation} \label{eq:cyclePow}
    \mathcal{P}_{\text{cyc}} = \mathcal{P}_{\text{lin}} - \frac{\mathcal{E}_{\text{cyc}}^2}{R_{\text{cyc}}} \ ,
\end{equation}
where $R_{\text{cyc}} \equiv \sum_{x=0}^n r(x)$ is the total resistance around the cycle, $\mathcal{E}_{\text{cyc}} \equiv \sum_{x=0}^n r(x) \dot{\Pi}(x)$ is the net ``electromotive force'' around the loop, and $\mathcal{P}_{\text{lin}}$ is the dissipated power for the linear graph $\curly{0,\dots,n}$ [\cref{eq:linPow}]. 

By expanding \eqref{eq:cyclePow} in terms of an arbitrary control set as in \eqref{eq:linPow}, we obtain $\tilde{\zeta}_{ij}^{\text{cyc}} = \tilde{\zeta}^{\text{lin}}_{ij} - \tilde{\zeta}^{\text{loop}}_{ij}$ where $\tilde{\zeta}^{\text{lin}}_{ij}$ is the friction for the linear graph and
\begin{equation}
    \tilde{\zeta}^{\text{loop}}_{ij} = \frac{\sum_{x,y}r(x)\,r(y)\, 
    \partial_i\Pi(x)\,\partial_j\Pi(y)}{\sum_x r(x)}
\end{equation}
is the reduction in the friction due to closure of the loop. 

The strict negativity of the correction $-\mathcal{E}_{\text{cyc}}^2/R_{\text{cyc}}$ to the linear-chain excess power in \cref{eq:cyclePow} reflects Rayleigh's monotonicity theorem: adding an edge to the graph can never increase effective resistances~\cite{doyle_random_2000}. Physically, the loop provides a parallel pathway that shunts probability flux, inherently reducing the overall thermodynamic cost.

For a distribution sufficiently localized away from the cut and slowly changing, the correction becomes negligible and the graph can effectively be treated as a linear graph. This follows from $R_{\text{cyc}} = \sum_x [w(x+1|x)\pi(x)]^{-1}$ and $\mathcal{E}_{\text{cyc}} = \sum_x \dot{\Pi}(x)
[w(x+1|x)\pi(x)]^{-1}$: If we cut an edge $(x,x+1)$ where $\pi(x), \pi(x+1)$, and their time derivatives are all very small, then $R_{\text{cyc}}$ will become very large while $\mathcal{E}_{\text{cyc}}$ remains bounded.

This method is generalizable: Decompose the total currents into a reference current on the same nodes and apply Thomson's principle to find the correction currents (which in general will not be spatially uniform). This could be applied, e.g., to determine the sensitivity of the LR excess power to changes in the topology of the Markov graph.

\sec{Continuous-state generalization}
The relationship between the LR dissipation and mean first-passage times established above for finite reversible Markov chains extends (with minor modifications) to continuous-space processes.

First, observe that for discrete state spaces the MFPT from $y$ to $x$ can be expressed as the integral
\begin{equation} \label{eq:mfpInt}
    \tau_{\rm{mfp}}(x|y) = \frac{1}{\pi(x)}\int_0^\infty \! \mathrm{d}t\, \sbrac{p_t(x|x) - p_t(x|y)} \ ,
\end{equation}
with $p_t(x|y)=\exp\curly{t\mathbb{W}}(x,y)$. This follows from \eqref{eq:DrazinMFPT} and the integral representation of the Drazin inverse of the rate matrix~\cite{coolen-schrijnerDeviationMatrixContinuoustime2002}.

We map this to a continuous state space $\Omega$ by replacing the discrete rate matrix $\mathbb{W}$ with a continuous infinitesimal generator (Fokker-Planck operator) $\mathscr{L}^\dagger$. Under detailed balance, the transition kernel $p_t(x|y) = \exp\curly{t\mathscr{L}^\dagger}(x,y)$ obeys $\pi(y)p_t(x|y) = \pi(x)p_t(y|x)$ for invariant density $\pi$. For diffusion in a confining potential, \cref{eq:mfpInt} (with $x,y$ now taken to be continuous variables) is precisely equal to the MFPT between points $x$ and $y$ for $\Omega = \mathbb{R}$ \cite{bicout_first_1997}. For higher dimensions, pointwise MFPTs diverge; however, under standard assumptions~\cite{wangGeneralizedInversesTheory2018, pavliotisStochasticProcessesApplications2014} the system relaxes exponentially to the steady-state density, so the integral in \eqref{eq:mfpInt} remains finite and serves as a well-defined physical timescale connecting points $x$ and $y$.

We define the commute-time kernel in terms of \eqref{eq:mfpInt} as in \cref{eq:commtime}. Then the metric equivalence $\beta g \overset{\Delta}{\sim}-\tfrac{1}{2}C$ holds for the continuous kernel (see the Supplemental Material~\cite{supplemental_key} for the complete proof).

\sec{Conclusion}
The geometry of dissipation in slowly driven Markov processes admits several representations, each offering different tools for interpretation and calculation. Through the graph Laplacian $L$, we have unified the friction metric with effective resistance, commute times, and discrete optimal transport restricted to paths of equilibrium distributions. 

Mapping the dynamical system onto a resistor network offers powerful tools for calculation and interpretation. Using standard methods from circuit theory, we derived exact friction metrics for linear and cyclic topologies. These results effectively demonstrate the more general observation that additional transition pathways (i.e., additional edges on the Markov graph) reduce LR thermodynamic cost via Rayleigh's monotonicity theorem. The mapping also leads to a direct probabilistic interpretation of LR dissipation. Simultaneously, the commute-time embedding provides intuition for the local geometry of the thermodynamic manifold and identifies bottlenecks as physical distances that require energy to traverse. 

These results suggest several interesting directions for future research. For continuous harmonic potentials, exact minimizers of the excess work (beyond linear response) can be obtained from LR optimal protocols via a counterdiabatic correction~\cite{zhongLinearResponseEquivalence2024}; though here we have extended the correspondence between LR control and OT, it remains an open question whether analogous corrections can be constructed for discrete graph dynamics. Further work might explore the metric structure on the edge space given control over non-conservative forces, leverage data-driven estimation of resistance metrics from simulation or experiment~\cite{noe_commute_2016} for complex systems, or examine the implications of commute-time and bottleneck inequalities for efficient driving.

\sec{Acknowledgements}
We thank Antonio Patr\'on Castro and W.\ Callum Wareham (Simon Fraser University, Department of Physics) for feedback on the manuscript, and also thank the anonymous reviewers, whose exceptionally meticulous reading and insightful comments substantially improved the manuscript. This work was supported by NSERC CGS Master's and Doctoral scholarships (J.R.S.), an NSERC Discovery Grant RGPIN-2020-04950 (D.A.S.), and a Tier-II Canada Research Chair CRC-2020-00098 (D.A.S.).

\nocite{maesLocalDetailedBalance2021}
\nocite{seabrookTutorialSpectralTheory2023}
\nocite{schnakenbergNetworkTheoryMicroscopic1976}
\nocite{dixit_inferring_2015}
\nocite{chowEntropyDissipationFokkerPlanck2018}
\nocite{yoshimuraHousekeepingExcessEntropy2023}
\bibliography{bibliography.bib}

\clearpage
\onecolumngrid 
\begin{center}
  \textbf{\large Supplemental Material: Thermodynamic geometry of friction on graphs: Resistance, commute times, and optimal transport}\\[.2cm]
  Jordan R.\ Sawchuk and David A.\ Sivak\\[.1cm]
  {\itshape Department of Physics, Simon Fraser University, Burnaby, British Columbia, Canada V5A1S6}
\end{center}
\vspace{1cm}

\setcounter{equation}{0}
\setcounter{figure}{0}
\setcounter{table}{0}
\setcounter{page}{1}
\setcounter{section}{0}

\renewcommand{\theequation}{S\arabic{equation}}
\renewcommand{\thefigure}{S\arabic{figure}}
\renewcommand{\thetable}{S\arabic{table}}
\renewcommand{\thesection}{S\arabic{section}}
\renewcommand{\theHequation}{S\arabic{equation}}
\renewcommand{\theHfigure}{S\arabic{figure}}
\renewcommand{\theHtable}{S\arabic{table}}
\renewcommand{\theHsection}{S\arabic{section}}

\section{Extended derivations}
\subsection{LR excess work}

We present here a sketch of the derivation of the linear response excess work. For an alternative treatment from the perspective of dynamical linear response theory, see~\cite{sivakThermodynamicMetricsOptimal2012} and its extension to full control in~\cite{sawchuk_global_2026}. We begin with a derivation of the linear response approximation of the lag $\delta \bm{p}_s$, roughly following Ref.~\cite{mandalAnalysisSlowTransitions2016}:
\begin{subequations}
    \begin{align}
        \frac{1}{\tau_{\text{prot}}}\frac{\partial \bm{p}_s}{\partial s} &= \frac{1}{\tau_{\text{prot}}}\frac{\partial \bm{\pi}_s}{\partial s} + \frac{1}{\tau_{\text{prot}}}\frac{\partial (\delta \bm{p}_s)}{\partial s} \\
        \mathbb{W}_s\delta \bm{p}_s &= \frac{1}{\tau_{\text{prot}}}\frac{\partial \bm{\pi}_s}{\partial s} + \frac{1}{\tau_{\text{prot}}}\frac{\partial( \delta \bm{p}_s)}{\partial s}   \label{eq:line2} \\
        \delta \bm{p}_s &= \frac{1}{\tau_{\text{prot}}}\mathbb{W}^{\mathcal{D}}_s \frac{\partial \bm{\pi}_s}{\partial s} +  \frac{1}{\tau_{\text{prot}}}\mathbb{W}^{\mathcal{D}}_s\frac{\partial (\delta \bm{p}_s)}{\partial s} \label{eq:line3}   \\ 
        &\approx \frac{1}{\tau_{\text{prot}}}\mathbb{W}^{\mathcal{D}}_s \frac{\partial \bm{\pi}_s}{\partial s} &(\tau_{\text{prot}} \text{ sufficiently long}) \label{eq:line4} 
    \end{align}
\end{subequations}
The substitution on the left-hand side of \eqref{eq:line2} follows from the master equation, and \eqref{eq:line3} follows from $\mathbb{W}^{\mathcal{D}}_s\mathbb{W}_s = I - \bm{\pi}_s\bm{1}^{\mathsf{T}}$. \Cref{eq:line3} holds exactly for any control protocol. In Ref.~\cite{mandalAnalysisSlowTransitions2016}, the approximation~\eqref{eq:line4} is recursively substituted into~\eqref{eq:line3} to generate a formal series expansion of $\delta \bm{p}_s$ in powers of $\tau_{\text{prot}}^{-1}$. Ref.~\cite{avron_adiabatic_2012} provides a mathematically rigorous treatment, addressing more directly the question of convergence. We direct the interested reader to those sources, offering here only a non-rigorous argument: The factors $\mathbb{W}^{\mathcal{D}}_s$ and $\dot{\bm{\pi}}_s$ in scaled time $s = t /\tau_{\text{prot}}$ characterize only the path through the probability simplex and the relative velocity from point to point, but not the absolute velocity. They are therefore independent of $\tau_{\text{prot}}$, so the first term on the right-hand side of~\eqref{eq:line3} is $\mathcal{O}(\tau_{\text{prot}}^{-1})$. Meanwhile, the second term on the right-hand side vanishes \textit{faster} than $\mathcal{O}(\tau_{\text{prot}}^{-1})$, since (given $\bm{p}_0 = \bm{\pi}_0$) $\delta \bm{p}_s \to 0$ in the quasistatic limit ($\tau_{\text{prot}} \to \infty$). Therefore, the first term is expected to dominate for sufficiently long $\tau_{\text{prot}}$.

We now use this approximation to derive the friction metric. Recalling that $\langle \mathcal{W}_{\text{ex}}\rangle = \int_0^1 \md s \,\dot{\bm{V}}^{\mathsf{T}}_s\delta \bm{p}_s$, we have (omitting the subscript $s$)
\begin{subequations}
    \begin{align}
        \dot{\bm{V}}^{\mathsf{T}}\delta \bm{p} &\approx \frac{1}{\tau_{\text{prot}}}\dot{\bm{V}}^{\mathsf{T}}\mathbb{W}^{\mathcal{D}}\dot{\bm{\pi}} \label{eq:2line1} \\
        &= \frac{1}{\tau_{\text{prot}}}\dot{\bm{V}}^{\mathsf{T}}\mathbb{W}^{\mathcal{D}}\frac{\partial \bm{\pi}}{\partial \bm{V}}\dot{\bm{V}} \label{eq:2line2} \\
        &= \frac{1}{\tau_{\text{prot}}}\dot{\bm{V}}^{\mathsf{T}}\paren{-\beta\mathbb{W}^{\mathcal{D}}D_{\bm{\pi}}}\dot{\bm{V}} \label{eq:2line3} \\
        &\equiv \frac{1}{\tau_{\text{prot}}}\dot{\bm{V}}^{\mathsf{T}}\zeta \, \dot{\bm{V}} \ .
    \end{align}
\end{subequations}
Eq.~\eqref{eq:2line3} follows from the fact that $\partial \bm{\pi}/\partial \bm{V} = \beta \paren{\bm{\pi}\bm{\pi}^{\mathsf{T}} - D_{\bm{\pi}}}$ for the Boltzmann distribution, together with $\mathbb{W}^{\mathcal{D}}\bm{\pi} = 0$.

For the corresponding equation on the simplex, observe that
\begin{subequations}
    \begin{align}
        -D_{\bm{\pi}}^{-1} \dot{\bm{\pi}} &=-D_{\bm{\pi}}^{-1}\frac{\partial \bm{\pi}}{\partial \bm{V}}\dot{\bm{V}} \\
        &= \beta (I - \bm{1}\bm{\pi}^{\mathsf{T}})\dot{\bm{V}}  \\
        &= \beta \,\dot{\bm{V}} - c \bm{1}  \ , 
    \end{align}
\end{subequations}
which follows from $\bm{1}^{\mathsf{T}}\mathbb{W}^{\mathcal{D}}=0$. We may then conclude that $\dot{\bm{\pi}}^{\mathsf{T}}\,g\,\dot{\bm{\pi}} = \dot{\bm{V}}^{\mathsf{T}}\zeta\, \dot{\bm{V}}$ 
with $g \equiv -k_{\rm B} T \,D_{\bm{\pi}}^{-1}\mathbb{W}^{\mathcal{D}}$.

\subsection{Proof that \texorpdfstring{$\beta g = (I - \bm{\pi}\bm{1}^{\mathsf{T}})^{\mathsf{T}}L^+(I - \bm{\pi}\bm{1}^{\mathsf{T}})$}{beta g = (I - pi 1^T)^T L^+ (I - pi 1^T)}}
\noindent Set $\beta = 1$. The matrices $g$ and $M \equiv (I - \bm{\pi}\bm{1}^{\mathsf{T}})^{\mathsf{T}}L^+(I - \bm{\pi}\bm{1}^{\mathsf{T}})$ are both symmetric, annihilate $\bm{\pi}$, and are right inverses of $L$ on the simplex:
\begin{subequations}
    \begin{align}
        Lg &= (\mathbb{W}D_{\bm{\pi}})(D_{\bm{\pi}}^{-1}\mathbb{W}^{\mathcal{D}}) \\ &= \mathbb{W} \mathbb{W}^{\mathcal{D}} \\ &= I - \bm{\pi}\bm{1}^{\mathsf{T}} 
    \end{align}
\end{subequations}
and
\begin{subequations}
    \begin{align}
        LM &= L(I - \bm{1}\bm{\pi}^{\mathsf{T}})L^+(I - \bm{\pi}\bm{1}^{\mathsf{T}}) \\ 
        &= LL^+ (I - \bm{\pi}\bm{1}^{\mathsf{T}}) \\ 
        &= (I - \frac{1}{|\Omega|}\bm{1}\bm{1}^{\mathsf{T}})(I - \bm{\pi}\bm{1}^{\mathsf{T}}) \\ 
        &=I - \bm{\pi}\bm{1}^{\mathsf{T}}
    \end{align}
\end{subequations}
Since the null space of $L$ is spanned by $\bm{1}$, and since both $g$ and $M$ are symmetric, $L(g - M) = 0 \implies g - M = c\bm{1}\bm{1}^{\mathsf{T}}$. Then since $g\bm{\pi}= M\bm{\pi} = 0$,
\begin{equation}
    0=(g-M)\bm{\pi} = c\bm{1}(\bm{1}^\mathsf{T}\bm{\pi}) = c \bm{1} \implies c = 0
\end{equation}
and therefore $g = M$. 

\section{Converting between metric representations}
Here we address inter-conversion between the equivalent metrics $\beta g, L^+, C,$ and $R_{\text{eff}}$. The metric equivalence $M\overset{\Delta^n}{\sim}M'$ means that $M,M'$ induce the same quadratic form over tangent vectors ($\dot{\bm{\pi}}$ such that $\bm{1}^{\mathsf{T}}\dot{\bm{\pi}}=0$). Then for $M,M'\in \mathbb{R}^{|\Omega|\times|\Omega|}$ symmetric, the relation is equivalently defined as
    \begin{equation} \label{eq:equivalt}
        M\overset{\Delta^n}{\sim} M' \iff M - M' = \bm{1}\bm{v}^{\mathsf{T}} + \bm{v} \bm{1}^{\mathsf{T}} 
    \end{equation}
    for $\bm{v} \in \mathbb{R}^{|\Omega|}$. The ``$\impliedby$'' direction follows immediately on contracting the right-hand side with tangent vectors. To show the ``$\implies$'' direction, let $D \equiv M - M'$, and define $P \equiv (I - \tfrac{1}{|\Omega|}\bm{1}\bm{1}^{\mathsf{T}})$, the projector onto the tangent space. Equality of the quadratic forms means that $PDP = 0$. One may then verify by direct substitution that $D = D - PDP$ is of the form $\bm{1}\bm{v}^{\mathsf{T}} + \bm{v}\bm{1}^{\mathsf{T}}$ with $\bm{v} = \frac{1}{|\Omega|}D\bm{1} - \frac{1}{2|\Omega|^2}(\bm{1}^{\mathsf{T}}D\bm{1})\bm{1}$. 

It follows from this definition~\cref{eq:equivalt} and $R_{\text{eff}} \overset{\Delta^n}{\sim} C$ that the term-wise equivalence
\begin{equation}
    R_{\text{eff}} = C
\end{equation}
holds exactly: the effective resistance between two states is identical to the commute time between them. This follows from $R_{\text{eff}}(x,x)= C(x,x)=0$, which forces $\bm{v} = 0$ in~\cref{eq:equivalt}. Note that this differs from the classical commute-time/resistance theorem of Ref.~\cite{chandra_electrical_1989}, which is defined in discrete time and carries a constant of proportionality.

The effective resistance (and therefore the commute-time matrix) can be obtained directly from the Laplacian pseudoinverse as $R_{\text{eff}}(x,y) = L^+(x,x)+L^+(y,y)-2L^+(x,y)$ [\cref{eq:effRes}], equivalently $R_{\text{eff}} = (\bm{1}\bm{d}^{\mathsf{T}} + \bm{d}\bm{1}^{\mathsf{T}}) - 2L^+$ for $\bm{d} \equiv \mathrm{diag}(L^+)$. Since $L^+$ is orthogonal to $\bm{1}$, 
\begin{equation} \label{eq:LplusRC}
    L^+ = -\tfrac{1}{2}P R_{\text{eff}} P = -\tfrac{1}{2}P C P 
\end{equation}
It is therefore also possible to obtain the edge conductances from the commute-time matrix by taking the Moore-Penrose pseudoinverse of the right-hand side of~\cref{eq:LplusRC}:
\begin{equation}
    w(x|y)\pi(y) = \sbrac{-\tfrac{1}{2}P C P }^+(x,y) \ .
\end{equation}
For detailed balance and a known equilibrium distribution $\bm{\pi}$, the commute times thus also uniquely determine the rate matrix $\mathbb{W}$. 

\Cref{eq:glap} gives the means to obtain $\beta g$ from $L^+$. The converse (obtaining $L^+$ from $\beta g$) again uses $L^+ \bm{1} = \bm{1}^{\mathsf{T}} L^+ = 0$, giving [as in~\cref{eq:LplusRC}]
\begin{equation}
    L^+ = P (\beta g) P \ .
\end{equation}

\color{black}
\section{Relaxing the detailed-balance condition \label{supp:rel}}
Throughout the main text, we assume global detailed balance. For more general systems obeying \textit{local} detailed balance
\begin{equation}
    -\ln\frac{w(x|y)}{w(y|x)} = \beta \sbrac{V(x) - V(y)} + \beta F(x,y) \ ,
\end{equation}
with non-conservative forces $F(x,y) = -F(y,x)$~\cite{maesLocalDetailedBalance2021}), the slow-driving/fast-relaxation asymptotic result $\delta \bm{p}^{\text{LR}} = \tau_{\text{prot}}^{-1}\mathbb{W}^{\mathcal{D}} \dot{\bm{\pi}}$ remains valid~\cite{avron_adiabatic_2012}. The core geometric structure therefore naturally extends to nonequilibrium systems, as shown in~\cite{mandalAnalysisSlowTransitions2016}. 

Here we demonstrate that the nonequilibrium extension is particularly sharp for conservative control [i.e., for fixed $F(x,y)$ and dynamically controlled $V(x)$] and transition rates of the Arrhenius form
\begin{equation} \label{eq:ArrRates}
    w(x|y) = \exp\curly{-\beta \sbrac{B(x,y) - V(y) +\tfrac{1}{2} F(x,y)}} \ ,
\end{equation}
where $B(x,y)=B(y,x)$ are symmetric activation barriers. For these rates, the stationary state $\bm{\pi}$ responds to changes in the potential exactly as the Boltzmann distribution does~\cite{owen_universal_2020}:
\begin{equation} \label{eq:responseBoltz}
    \frac{\partial \bm{\pi}}{\partial (\beta\bm{V})} = \bm{\pi}\bm{\pi}^{\mathsf{T}} - D_{\pi} \ .
\end{equation}
The metric tensor in this case is therefore
\begin{equation} \label{eq:nondbsymdef}
    \beta g \overset{\Delta^n}{\sim} \mathrm{Sym}\paren{-D_{\pi}^{-1}\mathbb{W}^{\mathcal{D}}} \ .
\end{equation}
where $\mathrm{Sym}(A) = \frac{1}{2}(A+A^{\mathsf{T}})$. For other rate laws, the response deviates from the Boltzmann form~\eqref{eq:responseBoltz}, inducing additional corrections to the metric.

\Cref{eq:nondbsymdef} immediately implies that $\beta g \overset{\Delta^n}{\sim} -\tfrac{1}{2}C$ for the rates~\eqref{eq:ArrRates}, since \cref{eq:DrazinMFPT} (connecting the Drazin inverse of the rate matrix to mean first-passage times) applies to all ergodic continuous-time Markov chains~\cite{coolen-schrijnerDeviationMatrixContinuoustime2002}. 

We capture the irreversibility of flow in a NESS by defining the forward and backward asymmetric Laplacians $L_{\text{fwd}} \equiv - \mathbb{W} \,D_{\pi}$ and $L_{\text{bwd}} \equiv -D_{\pi}\,\mathbb{W}^{\mathsf{T}}$, which naturally reduce to the standard symmetric Laplacian under detailed balance. Furthermore,
\begin{equation}
    \beta g \overset{\Delta^n}{\sim} \mathrm{Sym}(L_{\text{fwd}}^+) = \mathrm{Sym}(L^+_{\text{bwd}}) \ .
\end{equation}
The first equivalence holds for precisely the same reason as for the symmetric Laplacian discussed in the main text, and the second equivalence holds because $L_{\text{bwd}} = (L_{\text{fwd}})^{\mathsf{T}}$ and inversion commutes with transposition. Defining potentials $\phi$ through $\dot{\bm{\pi}} = L_{\text{fwd}}\phi$, their interpretation in the gauge $\bm{1}^{\mathsf{T}}\phi = 0$ is identical to the detailed-balance case:
\begin{equation}
    \phi(x) =\frac{1}{\tau_{\text{prot}}}\frac{\delta p^{\text{LR}}(x)}{\pi(x)} \ .
\end{equation}
Define the symmetric graph Laplacian $L \equiv \mathrm{Sym}(L_{\text{fwd}})$, with (negative) off-diagonal elements equal to half of the stationary traffic [equivalently, the average directed flux on the edge $(x,y)$]:
\begin{equation}
    -L(x,y) = \frac{1}{2}[w(x|y)\pi(y) + w(y|x)\pi(x)] \ .
\end{equation}
Defining the inner product on $\mathcal{H}(\mathcal{E})$ with respect to these symmetric weights gives $\langle \dot{\mathcal{W}}_{\text{ex}}\rangle^{\text{LR}} \propto  \phi^{\mathsf{T}}L\phi =\abs{\abs{\nabla \phi}}_{\mathcal{H}(\mathcal{E})}^2$. The equilibrium-restricted discrete Benamou-Brenier formula therefore extends to the nonequilibrium dynamics with continuity equation $\dot{\bm{\pi}} = L_{\text{fwd}}\phi$. 

The circuit-theoretic picture is also preserved by taking $r(x,y)^{-1} = - L(x,y)$:
\begin{equation}
    \langle \dot{\mathcal{W}}_{\text{ex}}^{\text{LR}}\rangle \propto \frac{1}{2}\sum_{x,y}r(x,y) \, i(x,y)^2 
\end{equation}
where $i(x,y) = [\phi(x)-\phi(y)]/r(x,y)$ as before. Now, the total current injection $I(x) = - \sum_yi(x,y) = (L\phi)(x)$ at the nodes consists of both the protocol-driven currents $\dot{\pi}(x)$ and a background stationary current due to the NESS flow. Write $L = L_{\text{fwd}} - \frac{1}{2}J$, where $J \equiv 2 \, \mathrm{Skew}(L_{\text{fwd}}) = L_{\text{fwd}}- L_{\text{bwd}}$ is the matrix of stationary currents with elements
\begin{equation}
    j(x,y) = w(x|y)\pi(y) - w(y|x)\pi(x) \ .
\end{equation}
Then the node-injection currents are
\begin{equation} \label{eq:fullCurrents}
    I(x) = \dot{\pi}(x) + \tfrac{1}{2}(J\phi)(x) \ ,
\end{equation}
which reduces to $I(x) = \dot{\pi}(x)$ under detailed balance.

We can compare dissipation in systems with and without stationary currents in the following way. Let $\mathbb{W}_{\text{db}}$ be the additive reversibilization~\cite{seabrookTutorialSpectralTheory2023} of the rate matrix, with rates
\begin{equation}
    w_{\text{db}}(x|y) = \frac{1}{2}\sbrac{1 + \e{A(x,y)}}w(x|y) \ ,
\end{equation}
where $A(x,y)= \ln[w(x|y)\pi(y) \, / \, w(y|x)\pi(x)]$ are the edge affinities~\cite{schnakenbergNetworkTheoryMicroscopic1976}. This essentially balances the flows over edges, resulting in dynamics with the same edge traffic as the original dynamics (i.e., $L$ is unchanged) but has no stationary currents ($J = 0$). The excess work in the original dynamics is
\begin{equation}
    \frac{\tau_{\text{prot}}}{k_{\rm B} T}\langle \dot{\mathcal{W}}_{\text{ex}}\rangle^{\text{LR}} = \dot{\bm{\pi}}^{\mathsf{T}}L^+\dot{\bm{\pi}} - \paren{\tfrac{1}{2}J\phi}^{\mathsf{T}}L^+\paren{\tfrac{1}{2}J\phi} \ ,
\end{equation}
and since the first term on the right-hand side is the dissipation for the detailed-balanced system,
\begin{equation}
     \langle \dot{\mathcal{W}}_{\text{ex}}\rangle^{\text{LR}} \leq  \langle \dot{\mathcal{W}}_{\text{ex}}\rangle^{\text{LR}}_{\text{db}} \ .
\end{equation}
This inequality implies that for the rate law~\eqref{eq:ArrRates}, background stationary currents actively assist in transporting probability mass without incurring additional linear-response work, regardless of the orientation of the background currents.

\subsection{Three-state cycle}
For a three-state cycle driven by fixed edge affinities $A(x,y)$ maintaining a stationary current $j_{\text{ss}}$ (\cref{fig:three_state_ness}), the dissipation is scaled down compared to the detailed-balance case: 
\begin{equation}
    \langle \dot{\mathcal{W}}_{\text{ex}}\rangle^{\text{LR}} = \alpha \langle \dot{\mathcal{W}}_{\text{ex}}\rangle_{\text{db}}^{\text{LR}} 
\end{equation}
with
\begin{equation}
    \alpha = \frac{R_{\text{cyc}}}{R_{\text{cyc}} + \frac{1}{4}j_{\text{ss}}^2\,r(0)r(1)r(2)} \ .
\end{equation}
Here $R_{\text{cyc}}$ and $r(x)$ are defined as in the main text.

\begin{figure}
    \centering
    \resizebox{0.20\linewidth}{!}{
    \begin{tikzpicture}
        [
        node_style/.style = {circle, draw=black, thick, minimum size=0.6cm, fill=caledon!20, inner sep=0pt, font=\bfseries},
        elec_node/.style = {circle, draw, fill=black, minimum size=3pt, inner sep=0pt},
        circuitikz/bipoles/length=0.8cm,
        circuitikz/resistors/scale=1,
        ]
        
        \def\R{1.6} 

        \node[node_style] (0) at (90:\R) {$0$};
        \node[node_style] (1) at (330:\R) {$1$};
        \node[node_style] (2) at (210:\R) {$2$};

        \draw[bend left = 10, line width = 1 pt] (0) to node[midway, right=2pt]{} (1);
        \draw[bend left = 10, line width = 1 pt] (1) to node[midway, right=2pt]{} (2);
        \draw[bend left = 10, line width = 1 pt] (2) to node[midway, right=2pt]{} (0);

        \draw[->, >=Latex, color=navyBlue, line width=1pt] (150:0.5) arc (150:-150:0.5) node[midway, left] {};

        \node[color=navyBlue] () at (0:0) {$j_{\text{ss}}$};
        
    \end{tikzpicture}
    }
    \caption{Three-state cycle with stationary current $j_{\text{ss}}$.}
    \label{fig:three_state_ness}
\end{figure}

More transparently, define the dimensionless measures of nonequilibrium driving $a_x \equiv \tanh \frac{1}{2}A(x,x+1)$, physically representing the stationary current divided by the total traffic over an edge. Then the scaling factor is 
\begin{equation}
    \alpha = \frac{a_0 + a_1 + a_2}{a_0 + a_1 + a_2 + a_0a_1a_2} \ .
\end{equation}
It is then straightforward to verify that $\frac{3}{4}< \alpha \leq 1$, where the lower bound is saturated in infinitely strong nonconservative driving [$A(x,y)\to \infty$]. 

Geometrically, the metrics $g$ and $g_{\text{db}}$ are conformally equivalent: Local angles between paths are exactly preserved, but infinitesimal distances are scaled by a factor $\sqrt{\alpha}$. Measured between common distributions, distances on the manifold $(\Delta^2,g)$ are strictly shorter than distances on the manifold $(\Delta^2, g_{\text{db}})$, but by no more than a factor $\sqrt{3}/2 \approx 0.87$.

\section{Discrete calculus and optimal transport: Formal definitions\label{supp:OTconn}}
\subsection{Discrete calculus}
Here we provide the formal definitions for discrete calculus used in the main text, following \cite{grady_discrete_2010} and later taking the conventions of \cite{maas_gradient_2011}. The need for a careful treatment can be seen in the expression $\rho_s\nabla\phi_s$ in the continuity equation~\eqref{eq:BBD}: because continuous vector fields map to edge functions and scalars to node functions, the product of a density and a gradient requires a formal definition to be mathematically well-posed.

For Markov graph $G = (\Omega,\mathcal{E})$, denote by $\mathcal{H}(\Omega)$ and $\mathcal{H}(\mathcal{E})$ the respective Hilbert spaces of vertex functions and edge functions. Analogous to their role in continuous calculus, the \textit{graph gradient} $\nabla_G:\mathcal{H}(\Omega) \to \mathcal{H}(\mathcal{E})$ and \textit{graph divergence} $\text{div}_G :\mathcal{H}(\mathcal{E}) \to \mathcal{H}(\Omega)$ map functions between these spaces. The inner products $\paren{\cdot,\cdot}_{\mathcal{H}(\Omega)}$ and $(\cdot,\cdot)_{\mathcal{H}(\mathcal{E})}$ on these spaces are required to obey an adjointness relation analogous to integration by parts and must reproduce the graph Laplacian~\eqref{eq:graphLap}:
\begin{equation} \label{eq:constraints}
    \begin{aligned}
        \paren{\nabla_G \varphi, \Psi}_{\mathcal{H}(\mathcal{E})} &= \paren{\varphi, -\text{div}_G\Psi}_{\mathcal{H}(\Omega)} \ , \\
        \varphi^{\mathsf{T}}L \,\psi &= \paren{\varphi,-\text{div}_G\nabla_G \psi}_{\mathcal{H}(\Omega)}  \ . 
    \end{aligned}
\end{equation}
These constraints do not uniquely determine the inner products and differential operators. Here we follow the conventions of~\cite{maas_gradient_2011}, defining the weighted inner products
\begin{subequations}
\begin{align}
(\psi,\varphi)_{\mathcal{H}(\Omega)} &= \sum_{x\in\Omega}\pi(x) \, \psi(x) \, \varphi(x) \\
(\Psi,\Phi)_{\mathcal{H}(\mathcal E)} &= \frac{1}{2}\sum_{x,y\in\Omega} w(x|y) \, \pi(y)\Psi(x,y)\,\Phi(x,y) \ , \label{eq:HEip}
\end{align}
\end{subequations}
and gradient and divergence operators
\begin{subequations}
    \begin{align}
        (\nabla_G\psi)(x,y) &\equiv \psi(x) - \psi(y) \\
        (\mathrm{div}_G \Psi)(x) &\equiv \frac{1}{2}\sum_{y \in \Omega}w(y|x) [\Psi(y,x) - \Psi(x,y)] \ .
    \end{align}
\end{subequations}

\subsection{Connections to previous work on discrete OT}
We show here that the restricted $L^2$-Wasserstein metric~\eqref{eq:equivOT} defined in the main text is a special case of the metric for probability transport on finite graphs~\cite{maas_gradient_2011, chowFokkerPlanckEquations2012,chowEntropyDissipationFokkerPlanck2018, liTransportInformationGeometry2022}. Consider a weighted graph $G=(\Omega,\mathcal{E},\omega)$ with vertex set $\Omega$, edge set $\mathcal{E}$, and symmetric edge weights $\omega(x,y)=\omega(y,x) > 0$ for $(x,y)\in\mathcal{E}$. The discrete $L^2$-Wasserstein distance between probability vectors $p_0, p_1$ on $G$ is defined in \cite{chowEntropyDissipationFokkerPlanck2018} as 
\begin{equation} \label{eq:chowW2}
    \mathcal{W}^2_2(p_0,p_1) \equiv \inf_{\dot{p}_s =- \text{div}_G(p_s \nabla_G \phi_s)}\int_0^1 \md s\abs{\abs{\nabla_G \phi_s}}^2_{p_s}   \ .
\end{equation}
The product $p \, \nabla_G \phi$ in the constraint is called a \textit{flux function}, defined as
\begin{equation}
    (p \, \nabla_G \phi)(x,y) 
    \equiv
    \theta_p(x,y)\, \nabla_G\phi(x,y)
\end{equation}
for some symmetric generalized mean $\theta_p(x,y)$ of $p(x)$ and $p(y)$, with divergence
\begin{equation}
    \text{div}_G(p\nabla_G \phi)(x) = -\sum_y \sqrt{\omega(x,y)}\,\theta_p(x,y)\,(\nabla_G\phi)(x,y)  \ .
\end{equation}
The gradient operator is $\sqrt{\omega}$-weighted,
\begin{equation}
    (\nabla_G \phi)(x,y) = \begin{cases}
        \sqrt{\omega(x,y)} \sbrac{\phi(x)-\phi(y)} 
        & (x,y)\in \mathcal{E}, \\
        0 & (x,y) \not\in\mathcal{E}
    \end{cases} \ ,
\end{equation}
and the inner product with respect to $p$ is 
\begin{equation}
    (v,u)_p 
    \equiv
    \frac{1}{2}\sum_{(x,y)\in\mathcal{E}}v(x,y)\theta_p(x,y)u(x,y)  \ .  
\end{equation}

Under the restriction $p_s = \pi_s$, and $\omega,\theta_{\pi}$ chosen such that 
\begin{equation} \label{eq:weightRestric}
    \omega(x,y)\,\theta_{\pi}(x,y) = w_{\pi}(x|y)\,\pi(y) \ ,
\end{equation}
the $L^2$-Wasserstein distance~\eqref{eq:chowW2} coincides exactly with the expression~\eqref{eq:equivOT} for the thermodynamic distance. Here we have emphasized in the notation $w_\pi(x|y)$ that the rates depend on the equilibrium distribution. 

The weights $\omega$ are $\pi$-independent and may refer to a fixed reference process. In the absence of physical motivation to the contrary, it is natural to take unit weights
\begin{equation}
    \omega(x,y) = \begin{dcases}
        1 \ , & (x,y) \in \mathcal{E} \\
        0 \ , & (x,y) \not\in \mathcal{E}
    \end{dcases} \ ,
\end{equation}
so that
\begin{equation}
    \theta_\pi(x,y) = w_{\pi}(x|y)\,\pi(y)  \ . 
\end{equation}
Under commonly chosen rate laws, $\theta_{\pi}(x,y)$ is indeed a generalized average. For instance, taking the rates $w_{\pi}(x|y) = \sqrt{\pi(x)/\pi(y)}$ that maximize trajectory entropy subject to detailed balance~\cite{dixit_inferring_2015} gives
\begin{equation}
    \theta_\pi(x,y) = \sqrt{\pi(x) \, \pi(y)} \ ,
\end{equation}
the geometric mean of the equilibrium probabilities. Glauber rates $w(x|y) = \pi(x)/[\pi(x) +\pi(y)]$ give 
\begin{equation}
    \theta_\pi(x,y) = \sbrac{\frac{1}{\pi(x)}+\frac{1}{\pi(y)}}^{-1} \ ,
\end{equation}
the harmonic mean of the equilibrium probabilities. In \cite{maas_gradient_2011,chowFokkerPlanckEquations2012,chowEntropyDissipationFokkerPlanck2018}, the generalized average $\theta_p(x,y)$ is chosen such that the dynamics are a gradient flow with respect to some entropy or free-energy functional. Though it is not clear whether such gradient-flow structures are relevant in this context, the forms of $\theta_p$ studied in \cite{maas_gradient_2011,chowFokkerPlanckEquations2012,chowEntropyDissipationFokkerPlanck2018} can be reproduced with suitable transition rates.

Lastly, we briefly discuss the far-from-equilibrium discrete OT formulation presented by Yoshimura, et al.~\cite{yoshimuraHousekeepingExcessEntropy2023}. As in our construction, the Wasserstein distance between distributions in~\cite{yoshimuraHousekeepingExcessEntropy2023} is the infimum of an integral over the squared norm of a graph gradient on a weighted edge space:
\begin{equation} \label{eq:w2yoshi}
    \mathcal{W}_*(\bm{p}_0,\bm{p}_1)^2 = \inf_{\bm{p},\psi}\sbrac{\int_0^1 \md s \abs{\abs{\nabla_G \psi_s}}_{\mathcal{H}_*(\mathcal{E})}^2} 
\end{equation}
with $\psi$ defined by $\dot{\bm{p}} = L_*(\bm{p})\psi$ and $(\nabla_G\psi)(x,y)=\psi(x)-\psi(y)$, and asterisks here distinguish related objects in the two formulations. This similarity may reflect that the two constructions are both descendants of Maas' formulation~\cite{maas_gradient_2011} (a similar form is also presented in~\cite{chowEntropyDissipationFokkerPlanck2018}). However, they generalize in nearly orthogonal directions: \cref{eq:w2yoshi} allows $\bm{p}_t$ arbitrarily far from stationarity but holds the dynamics time-homogeneous, while our framework applies strictly close to stationarity but with time-inhomogeneous dynamics. Interestingly, the edge weights of Ref.~\cite{yoshimuraHousekeepingExcessEntropy2023} are the \textit{logarithmic mean} of the (non-stationary) forward and reverse fluxes on an edge, while our edge weights are the \textit{arithmetic mean} of the (stationary) forward and reverse fluxes on an edge (i.e., half the traffic). However, while defining the weights in this way holds for \textit{any} rate law in the genuinely nonequilibrium formalism of~\cite{yoshimuraHousekeepingExcessEntropy2023}, in our formalism it does not survive generalization beyond the Arrhenius-like rate law discussed in Sec.~\ref{supp:rel}, due to additional corrections to the metric.

\section{Commute-time kernel \label{supp:commtimekern}}
We show here that a \textit{commute-time kernel} $C(x,y)$ introduced in the main text by analogy to the discrete commute-time matrix is metrically equivalent to the friction tensor for continuous systems. Let $p_t(x|y) = \exp\curly{t\mathscr{L}^\dagger}(x,y)$ be the transition kernel of a continuous-space reversible Markov process with infinitesimal generator $\mathscr{L}^\dagger$. Define the \textit{commute-time kernel}
\begin{equation} \label{eq:commKern}
    C(x,y) \equiv \int_0^\infty \md t \ \sbrac{\frac{p_t(x|x) - p_t(x|y)}{\pi(x)} +\frac{p_t(y|y)-p_t(y|x)}{\pi(y)}} \ .
\end{equation}
As discussed in the main text, for $\Omega = \mathbb{R}$ this coincides with the actual commute time between points $x$ and $y$. For $\Omega = \mathbb{R}^d$ with $d > 1$, the interpretation is less straightforward, though $C(x,y)$ still describes a timescale connecting points $x$ and $y$. 

Let $\mathcal{A}$ and $\mathcal{B}$ be real-valued functions on $\Omega$ (i.e., observables) such that $\angbr{\mathcal{A}}_{\pi} = \angbr{\mathcal{B}}_{\pi} =0$. Then from the definition~\eqref{eq:commKern}, 
\begin{align}
\label{eq:gkr}
    \int_0^{\infty}\md t &\, \angbr{\mathcal{A}(X_t)\,\mathcal{B}(X_0)}_{\text{eq}} \\
    &= -\frac{1}{2}\int_{\Omega} \md x \int_{\Omega}\md y \ \pi(x)\,\mathcal{A}(x)\,C(x,y)\,\mathcal{B}(y)\,\pi(y) \ . \nonumber
\end{align}
[The $-1/2$ factor comes from the detailed-balance symmetry of the factors $-p_t(x|y)/\pi(x)$ and $-p_t(y|x)/\pi(y)$ in the integrand of $C(x,y)$.] Let $\mathcal{A}(X_t) = \hat{\omega}(x',X_t)$ and $\mathcal{B}(X_t) = \hat{\omega}(x'', X_t)$ be the relative empirical density fluctuations 
\begin{equation}
    \hat{\omega}(x,X_t) \equiv \frac{\delta(x-X_t) - \pi(x)}{\pi(x)} 
\end{equation}
at some fixed points $x',x''\in\Omega$. Then
\begin{subequations}
    \begin{align}
        \int_0^{\infty}\md t \, \big\langle \mathcal{A}(&X_t)\,\mathcal{B}(X_0)\big\rangle_{\text{eq}} \nonumber \\
        &= \int_0^{\infty}\md t \, \angbr{\hat{\omega}(x',X_t)\,\hat{\omega}(x'', X_t)}_{\text{eq}} \\
        &= k_{\rm B} T \frac{1}{\pi(x')\,\pi(x'')} \, \zeta(x',x'') \\
        &= \beta \, g(x', x'') \ , \label{eq:lastlinekernel}
    \end{align}
\end{subequations}
where $\zeta(x',x'')$ is the integral kernel of the continuous energy-space friction tensor~\cite{sawchuk_global_2026} and the final step follows from the change-of-variables formula
\begin{equation}
    \zeta(x',x'') = \int\md y'\int\md y'' \,\frac{\delta \pi(y')}{\delta V(x')}\,\frac{\delta \pi(y'')}{\delta V(x'')}\,g(y',y'') \ .
\end{equation} 

Next, substituting $\mathcal{A}(X_t)$ and $\mathcal{B}(X_t)$ into the right-hand side of \eqref{eq:gkr} gives
\begin{subequations} \label{eq:contmet}
    \begin{align}
        -\tfrac{1}{2}\int_{\Omega} \md x \int_{\Omega}\md y \ \pi(x)\,\mathcal{A}(x)\,C(x,y)\,\mathcal{B}(y)\,\pi(y)  &= -\tfrac{1}{2}\int_{\Omega}\md x \int_{\Omega}\md y \,C(x,y)\sbrac{\delta(x-x')-\pi(x')}\sbrac{\delta(y-x'')-\pi(x'')} \\
        &= - \tfrac{1}{2}C(x',x'') + k_1 \pi(x') + k_2 \pi(x'') \ ,
    \end{align}
\end{subequations}
for constants $k_1, k_2$. 
Since $\int_{\Omega}\md x \, \dot{\pi}(x)=0$, 
these constant-coefficient terms vanish in the LR excess power
\begin{equation}
    \angbr{\mathcal{P}_{\text{ex}}}^{\text{LR}} = \int_{\Omega}\md x\int_{\Omega}\md y \, \, g(x,y)\,\dot{\pi}(x)\,\dot{\pi}(y) \ ,
\end{equation}
and thus $\beta g \overset{\Delta}{\sim} -\tfrac{1}{2}C$.

\end{document}